\documentclass[journal,comsoc]{IEEEtran}
\usepackage[T1]{fontenc}
\usepackage{amsmath}
\usepackage[cmintegrals]{newtxmath}
\usepackage{graphicx}
\usepackage{epsfig}
\usepackage{amsfonts}
\usepackage{amssymb}
\usepackage{gensymb}
\usepackage{caption}
\usepackage{epstopdf}
\usepackage{tikz}
\usepackage{color}
\usepackage{array}
\usepackage{tabu}
\usepackage{algorithm}
\usepackage{subcaption}
\usepackage{relsize}
\usepackage{adjustbox}

\newcommand{\argmin}{\mathop{\text{argmin}}}

\begin{document}
\title{{\LARGE
A DNN Architecture for the Detection of Generalized \\ \vspace{-2mm}
Spatial Modulation Signals}}
\author{Bharath Shamasundar and A. Chockalingam \\
Department of ECE, Indian Institute of Science, Bangalore 560012
\vspace{-4mm}
}

\maketitle

\begin{abstract}
In this letter, we consider the problem of signal detection in generalized 
spatial modulation (GSM) using deep neural networks (DNN). 
We propose a novel modularized DNN architecture that uses 
small sub-DNNs to detect the active antennas and complex modulation symbols, 
instead of using a single large DNN to jointly detect the active antennas 
and modulation symbols. The main idea is that using small sub-DNNs instead 
of a single large DNN reduces the required size of the NN and hence requires 
learning lesser number of parameters. Under the assumption of i.i.d 
Gaussian noise, the proposed DNN detector achieves a performance very close 
to that of the maximum likelihood detector. We also analyze the performance 
of the proposed detector under two practical conditions: $i)$ correlated 
noise across receive antennas and $ii)$ noise distribution deviating from 
the standard Gaussian model. The proposed DNN-based detector learns the 
deviations from the standard model and achieves superior performance 
compared to that of the conventional maximum likelihood detector.
\end{abstract}

\pagestyle{empty}
\section{Introduction}
\label{Sec1}
Index modulation (IM) techniques are attracting increased research 
attention due to their superior bit error performance at lesser hardware 
complexity \cite{im1}. Spatial modulation (SM) \cite{sm1}-\cite{sm3} is 
a popular IM scheme which uses $n_t$ transmit antennas and a single 
transmit radio frequency (RF) chain. In a given channel use, one of the 
transmit antennas is selected based on $\lfloor\log_2n_t\rfloor$ 
information bits and a symbol from a modulation alphabet $\mathbb{A}$ 
(QAM/PSK) is transmitted on the selected antenna. Thus SM achieves a rate of 
$\lfloor\log_2n_t\rfloor + \log_2|\mathbb{A}|$ bits per channel use (bpcu). 
The reduced hardware complexity in SM comes at the cost of the reduced 
throughput. This drawback is overcome by generalized SM (GSM), which 
allows multiple transmit antennas to be active simultaneously 
\cite{gsm},\cite{gsm2}. GSM uses $n_t$ transmit antennas 
and $n_{rf}$ RF chains, $1<n_{rf} < n_t$. In each channel use, $n_{rf}$ 
out of the $n_t$ transmit antennas are selected based on 
$\lfloor \log_2 \binom{n_t}{n_{rf}} \rfloor$ information bits and 
$n_{rf}$ symbols from the modulation alphabet $\mathbb{A}$ are transmitted 
from the selected active antennas. The achieved rate in GSM is therefore 
$\lfloor \log_2 \binom{n_t}{n_{rf}} \rfloor + n_{rf}\log_2|\mathbb{A}|$ 
bpcu. In the present work, we consider the problem of signal detection 
for GSM using deep neural networks (DNN).

Recently, deep learning (DL) has been employed in wireless communications 
for designing intelligent communication systems 
\cite{DLforComm1}-\cite{DLsigdet}. Specifically, in the physical layer, 
DL has been applied in two important ways: $i)$ as a replacement to the 
existing communication blocks like channel coding \cite{DLforComm5} and 
signal detection \cite{DLsigdet}, \cite{dnn-tabu}, and $ii)$ for designing 
end-to-end communication systems without traditional communication blocks 
\cite{DLforComm4}. Both the approaches have shown promising results. 
DL has been applied in the context of SM in \cite{DNN_SM} 
to achieve link adaptation, in which the problems of transmit antenna 
selection (TAS) and power allocation (PA) are converted to those of 
data driven prediction, which are then solved using DNN-based methods. 
In the present work, we consider the problem of signal detection in GSM 
and explore the utility of DNN for detection task.  Our contributions in 
this letter can be summarized as follows.
\begin{itemize}
\item We propose a novel modularized DNN architecture that 
uses small sub-DNNs to detect the active antennas and complex modulation 
symbols. This is in contrast to using a single large DNN to jointly detect 
the active antennas and modulation symbols. The main idea is that using 
small sub-DNNs reduces the required size of the NN and hence requires 
learning lesser number of parameters.
\item We show that, under static channel conditions and i.i.d Gaussian 
noise across receive antennas, the proposed DNN architecture can achieve 
a performance very close to that of the optimum maximum likelihood detection. 
\item When the noise across different receive antennas are correlated, 
which arises in practice due to mutual coupling among the receive antennas,
matching networks, etc., the DNN-based detector learns the 
noise correlation and achieves superior performance compared to that of
using the maximum likelihood (ML) detection meant for i.i.d Gaussian noise, 
and a performance close to that of the true ML detector for correlated 
noise (which achieves the best detection performance under correlated 
noise). Also, when the noise is i.i.d but the distribution slightly 
deviates from Gaussian, the proposed DNN architecture learns a good 
detector for the non-Gaussian noise, and achieves superior 
performance compared to the ML detector meant for i.i.d Gaussian noise.
\item Finally, we extend the proposed DNN-based detector 
to the case of varying channels and show that the proposed detector 
achieves a performance close to that of the optimum ML detector. 
\end{itemize}
We note that, although low-complexity signal detection in GSM has been 
previously studied in the literature (e.g., \cite{gsm2},\cite{cs_det}), 
these works mainly consider GSM signal detection in the standard i.i.d 
Gaussian noise settings. GSM detectors under non-standard noise settings 
have not been reported, and the proposed DNN approach that considers 
GSM detection in non-standard noise settings is a novel contribution.

\section{GSM system model}
\label{Sec2}
Consider a MIMO communication system with $n_t$ transmit and $n_r$ receive 
antennas. Let $n_{rf}$, $1 < n_{rf} < n_t$, be the number of transmit RF 
chains at the transmitter. In GSM, in a channel use, $n_{rf}$ out of the 
$n_t$ transmit antennas are selected based on  
$\lfloor \log_2 \binom{n_t}{n_{rf}} \rfloor$ information bits. The selected 
$n_{rf}$ antennas are called active antennas, on which $n_{rf}$ symbols 
from a modulation alphabet $\mathbb{A}$ (say, QAM) are transmitted based 
on $n_{rf}\log_2|\mathbb{A}|$ information bits. Let 
$\mathbb{A}_0 = \mathbb{A} \cup 0$. The GSM signal set is a set of 
$n_t$-length vectors given by
\begin{equation}
\mathbb{S} = \{\mathbf{x}| \mathbf{x} \in \mathbb{A}_0, \|\mathbf{x}\|_0 = n_{rf}, \mathbf{t}^{\mathbf{x}} \in \mathbb{T}_A\},
\end{equation}
where $\mathbf{t}^{\mathbf{x}}$ is the 
antenna activation pattern (AAP) for the GSM signal vector $\mathbf{x}$ 
which is an $n_t$-length binary vector with 
$\mathbf{t}^{\mathbf{x}}_i=1$ if $\mathbf{x}_i \in \mathbb{A}$ and `0' 
otherwise, and $\mathbb{T}_A$ is the set of all valid AAPs. Denoting 
$\mathbf{H}$ to be the $n_r\times n_t$ MIMO channel matrix, the 
$n_r \times 1$ received signal vector $\mathbf{y}$ is given by 
\begin{equation}
\mathbf{y} = \mathbf{Hx} + \mathbf{n},
\label{eq:sys_model}
\end{equation} 
where $\mathbf{x} \in \mathbb{S}$ and $\mathbf{n}$ is an $n_r \times 1$ 
noise vector. Assuming perfect channel knowledge at the receiver, the 
maximum likelihood (ML) detection rule for GSM signal detection is given by
\begin{equation}
\hat{\mathbf{x}}=\argmin_{\mathbf{x}\in \mathbb{S}}\|\mathbf{y}-\mathbf{Hx}\|^2.
\label{eq:MLdet}
\end{equation} 
The ML detection 
rule in \eqref{eq:MLdet} is optimal only when the noise samples across the 
receive antennas are i.i.d and follow Gaussian distribution. Any deviation 
in noise from this standard model will result in suboptimal performance 
when \eqref{eq:MLdet} is used. This key observation motivates the use of 
DL techniques when there is deviation from the standard model. Accordingly, 
in the following sections, we propose a DNN architecture for GSM signal 
detection and assess its performance.

\begin{figure}
\centering
\includegraphics[height=5.5cm, width=8.25cm]{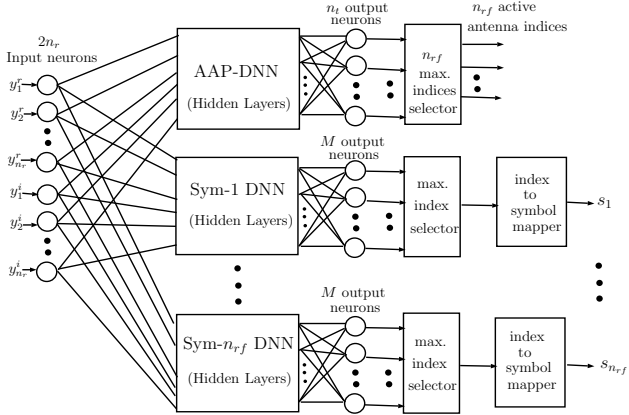}
\caption{Proposed DNN architecture for GSM signal detection.}
\vspace{-2mm}
\label{DGSM_blk_diag}
\end{figure}

\section{DNN-based GSM detector}
\label{Sec3}
GSM signal detection involves $i)$ detecting the set of $n_{rf}$ active 
antennas and $ii)$ detecting $n_{rf}$ modulation symbols 
$s_1,s_2,\cdots,s_{n_{rf}} \in \mathbb{A}$ transmitted from the active 
antennas. To do this, we propose the DNN architecture shown in 
Fig. \ref{DGSM_blk_diag}, which comprises of $n_{rf}+1$ smaller sub-DNNs. 
One sub-DNN is used to detect the indices of the $n_{rf}$ active antennas 
(which is shown as AAP-DNN) and $n_{rf}$ sub-DNNs are used for detecting 
$n_{rf}$ modulation symbols transmitted from the active antennas (which 
are shown as Sym-1 DNN, $\cdots$, Sym-$n_{rf}$ DNN). All the sub-DNNs 
have $2n_r$ input neurons through which the real and imaginary parts of 
the received signal vector are fed as inputs. 

{\em AAP-DNN:}
The AAP-DNN has a set of hidden layers and an output layer with $n_t$ 
neurons. Each neuron in the output layer corresponds to one transmit 
antenna and gives the probability of that antenna being active. We use 
sigmoid activation in the output layer so that the probabilities are 
independent across the output neurons and need not sum to one. 
The `$n_{rf}$ max. indices selector' takes the $n_t$ probability values 
from the output neurons as input and declares the $n_{rf}$ antennas 
corresponding to the  $n_{rf}$ highest probability values to be active. 

{\em Symbol-DNN:} 
Each of the Sym-$i$, $i=1,\cdots, n_{rf}$ DNNs has a set of hidden layers 
and $M=|\mathbb{A}|$ output neurons. Each output neuron of the Sym-$i$ DNN 
corresponds to one symbol of $\mathbb{A}$ and gives the probability of 
that symbol being sent from the $i$th active antenna.  Softmax activation 
is used for the output neurons of the symbol-DNNs. Hence, the probabilities 
in a given symbol-DNN are dependent across the output neurons and  sum 
to one. Only one of the $M$ neurons in each symbol-DNN will result 
in a high probability value, which will be declared as the transmitted symbol 
by the `max. index selector' followed by the `index to symbol mapper' blocks. 

A key advantage of the proposed DNN-based detector is that 
it has a modular architecture where the GSM signals are detected using 
small sub-DNNs instead of one large DNN. For example, consider a GSM system 
with $n_t=10$, $n_{rf}=4$, and 4-QAM. The signal set for this GSM system 
consists of 
{\small $2^{\lfloor\log_2\binom{10}{4}\rfloor + 4\log_2 4}= 2^{15}= 32768$}
signal vectors. It is known from the DL literature that using one-hot 
encoding for classification leads to excellent performance. For the 
considered GSM system, using a single DNN to achieve signal detection 
with one-hot encoding requires using 32768 output neurons. Further, the 
required number of hidden layers and the number of neurons in each hidden 
layer scale in proportion to the number of neurons in the output layer. 
A higher number of layers and a large number of neurons in each layer 
requires learning a large number of parameters during the training phase. 
The testing (signal detection) phase also gets complicated proportionately.
On the other hand, the proposed modular architecture for the considered 
GSM system uses five small sub-DNNs, {\em viz}., one AAP-DNN and four 
Symbol-DNNs. The AAP-DNN requires $n_t=10$ output neurons and the 
Symbol-DNNs require $|\mathbb{A}|=4$ output neurons. Therefore, compared 
to using a single large DNN, using small sub-DNNs requires a lower number 
of output neurons for each sub-DNN which, in turn, reduces the required 
number of hidden layers and the number of neurons in each hidden layer. 
Therefore, the training phase requires learning a lesser number of 
parameters and the testing phase is also simplified compared to using 
one large DNN.

{\em Training and Testing:} We consider a static/slowly varying 
channel with a long coherence-time so that the detector can be trained 
initially with $m_{T}$ labeled training examples and then subsequently be 
used for signal detection. In the training phase, the transmitter sends 
$m_{T}$ pseudo-random GSM signal vectors  known at both the transmitter 
and the receiver so that they can be used as labels for training the DNN. 
The received signal vectors generated according to the system model in 
\eqref{eq:sys_model} are used as inputs to train the AAP-DNN and 
symbol-DNNs. The number of training examples $m_T$ is selected based on 
experimentation where an initial $m_T$ of 1000 is used and is then 
increased gradually in steps of 1000 till a good classification (signal 
detection) performance is achieved. After the training phase, GSM signal 
vectors selected by the random information bits are transmitted in the 
testing phase and are detected using the trained DNN. We note that, 
during the training phase, the DNN learns the mapping from the received 
signal vectors to the transmitted GSM vectors, which is nothing but 
learning an equalizer for the channel corresponding to that coherence 
interval. Since the channel is static/slowly varying, the trained DNN 
can be used for several channel uses for signal detection until the 
channel is changed. Therefore, channel need not be explicitly made 
known at the receiver in the training process. TensorFlow and Keras 
framework are used for training and testing the proposed DNN architecture. 

\begin{table}[t]
\centering
\begin{tabular} {|l|c|c|}
\hline
{\footnotesize Parameters} & {\footnotesize APP-DNN} & {\footnotesize Symbol-DNN} \\ \hline \hline 
{\footnotesize No. of input neurons} & {\footnotesize $2n_r=8$} & {\footnotesize $2n_r=8$} \\ \hline 
{\footnotesize No. of output neurons} & {\footnotesize $n_t=4$} & {\footnotesize $|\mathbb{A}|=2$} \\ \hline 
{\footnotesize No. of hidden layers} & {\footnotesize 3} & {\footnotesize 3} \\ \hline  
{\footnotesize Hidden layer activation} & {\footnotesize ReLU} & {\footnotesize ReLU} \\ \hline
{\footnotesize Output layer activation} & {\footnotesize Sigmoid} & {\footnotesize Softmax} \\ \hline
{\footnotesize Optimizer} & {\footnotesize Adam} & {\footnotesize Adam} \\ \hline
{\footnotesize Loss function} & {\footnotesize Binary cross entropy} & {\footnotesize Binary cross entropy} \\ \hline
{\footnotesize No. of training examples} & {\scriptsize 10,000} & {\footnotesize 10,000} \\ \hline
{\footnotesize Training SNR} & {\footnotesize 10 dB} & {\footnotesize 10 dB} \\ \hline
{\footnotesize No. of epochs} & {\footnotesize 20} & {\footnotesize 20} \\ \hline   
\end{tabular} 
\caption{{DNN parameters of proposed detector in 
Fig. \ref{fig:ber_iid_awgn_small}.}}
\label{table1} 
\vspace{-2mm}
\end{table}

\section{Results and discussions}
\label{Sec4}
\subsection{BER with i.i.d Gaussian noise}
\begin{figure}[t]
\centering
\includegraphics[width=8cm, height=5.5cm]{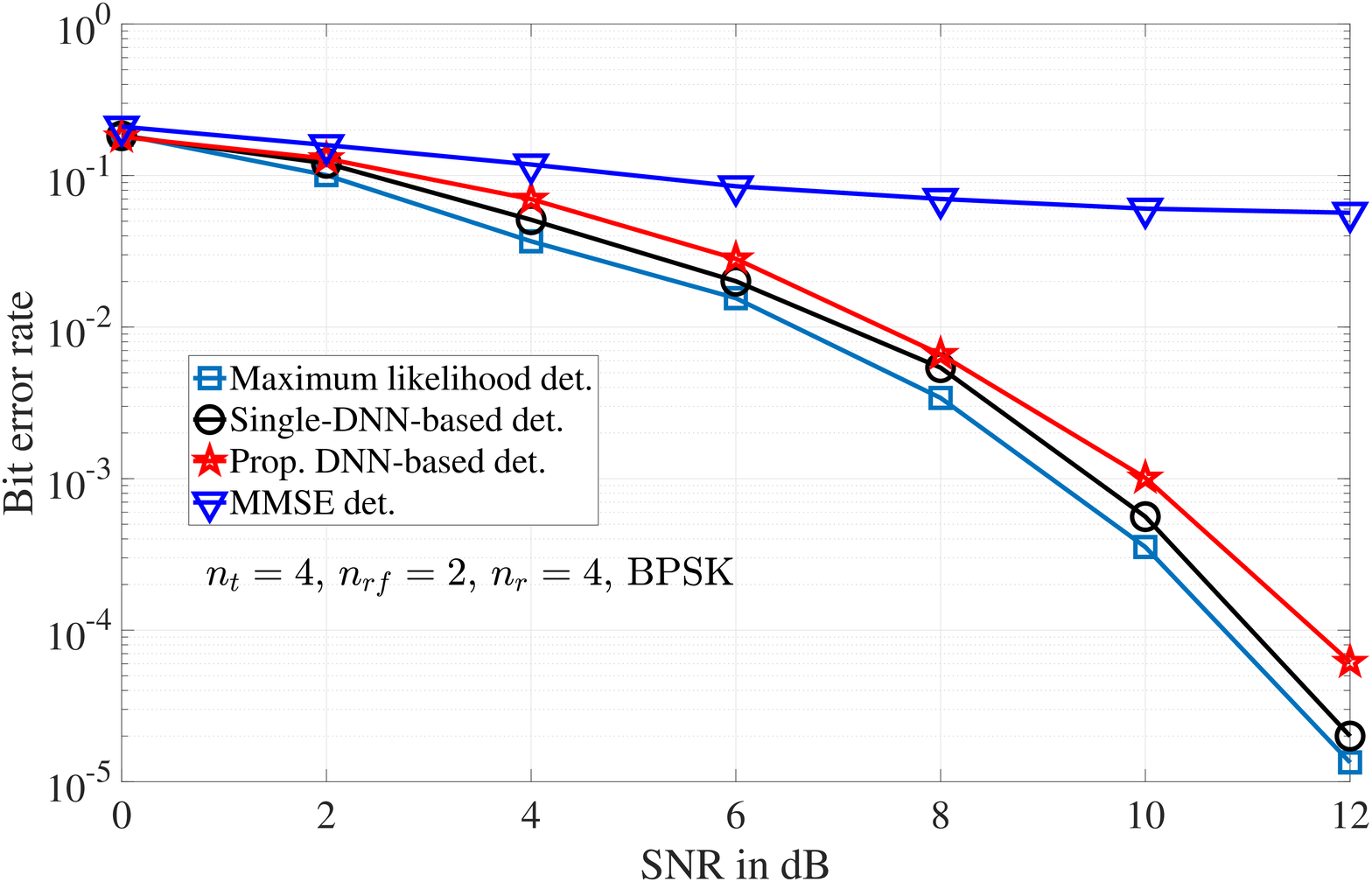}
\caption{BER performance of GSM with the proposed modularized DNN-based 
detector and single-DNN-based detector.}
\vspace{-2mm}
\label{fig:ber_iid_awgn_small}
\end{figure}

Figure \ref{fig:ber_iid_awgn_small} shows 
the BER performance of GSM using the proposed DNN-based detector, 
ML detector, MMSE detector, and a single-DNN-based detector 
(which uses a single large DNN to jointly detect active antenna indices 
and modulation symbols). The proposed DNN 
architecture has three sub-DNNs for this system setting, one sub-DNN 
for detecting the two active antennas and the other two sub-DNNs for 
detecting the two BPSK symbols transmitted by the active antennas. The 
DNN parameters used for the system configuration in Fig. 
\ref{fig:ber_iid_awgn_small} are presented in Table \ref{table1}.   
The following architectures are used: 
{\em {\bfseries APP-DNN:}}
$\text{i/p} \rightarrow 16\rightarrow \text{ReLU} \rightarrow 16 \rightarrow \text{ReLU} \rightarrow 8\rightarrow \text{ReLU} \rightarrow 4 \rightarrow \text{sigmoid}$. \\
{\em {\bfseries Symbol-DNN:}}
$\text{i/p} \rightarrow 16\rightarrow \text{ReLU} \rightarrow 16 \rightarrow \text{ReLU} \rightarrow 8\rightarrow \text{ReLU} \rightarrow 2 \rightarrow \text{softmax}$. \\
In the above architectures, numbers denote the number of neurons in a given 
layer, which is followed by the activation function used in that layer. 
The single-DNN-based detector architecture used is: \\
{\small 
{\em {\bfseries Single-DNN:}}
{\small i/p $\rightarrow 32 \rightarrow$ ReLu $\rightarrow 32 \rightarrow$ ReLu $\rightarrow 64 \rightarrow$ ReLu $\rightarrow 64 \rightarrow$ ReLu $\rightarrow 32 \rightarrow$ ReLu $\rightarrow 16 \rightarrow$ Softmax}.}  \\
The channel is considered to be static with the channel gains taking values 
from an instance of Rayleigh flat fading channel. From 
Fig. \ref{fig:ber_iid_awgn_small}, it is seen that the performance of the 
considered GSM system with the proposed detector is very close to that 
with the ML detector and is much superior to that with MMSE detector. 
The single-DNN-based detector performs slightly better than the proposed 
detector but this comes at the cost of significantly high complexity, 
e.g., the number of parameters to be learnt in the considered 
single-DNN-based detector and the proposed detector are 8128 and 1600, 
respectively. Further, the performance of single-DNN-based detector gets 
worse and inferior if fewer layers (and hence fewer parameters are to be 
learnt) are used.

\begin{figure}[t]
\centering
\begin{subfigure}[b]{0.25\textwidth}
\hspace*{-4mm}
\includegraphics[width=5 cm, height=4 cm]{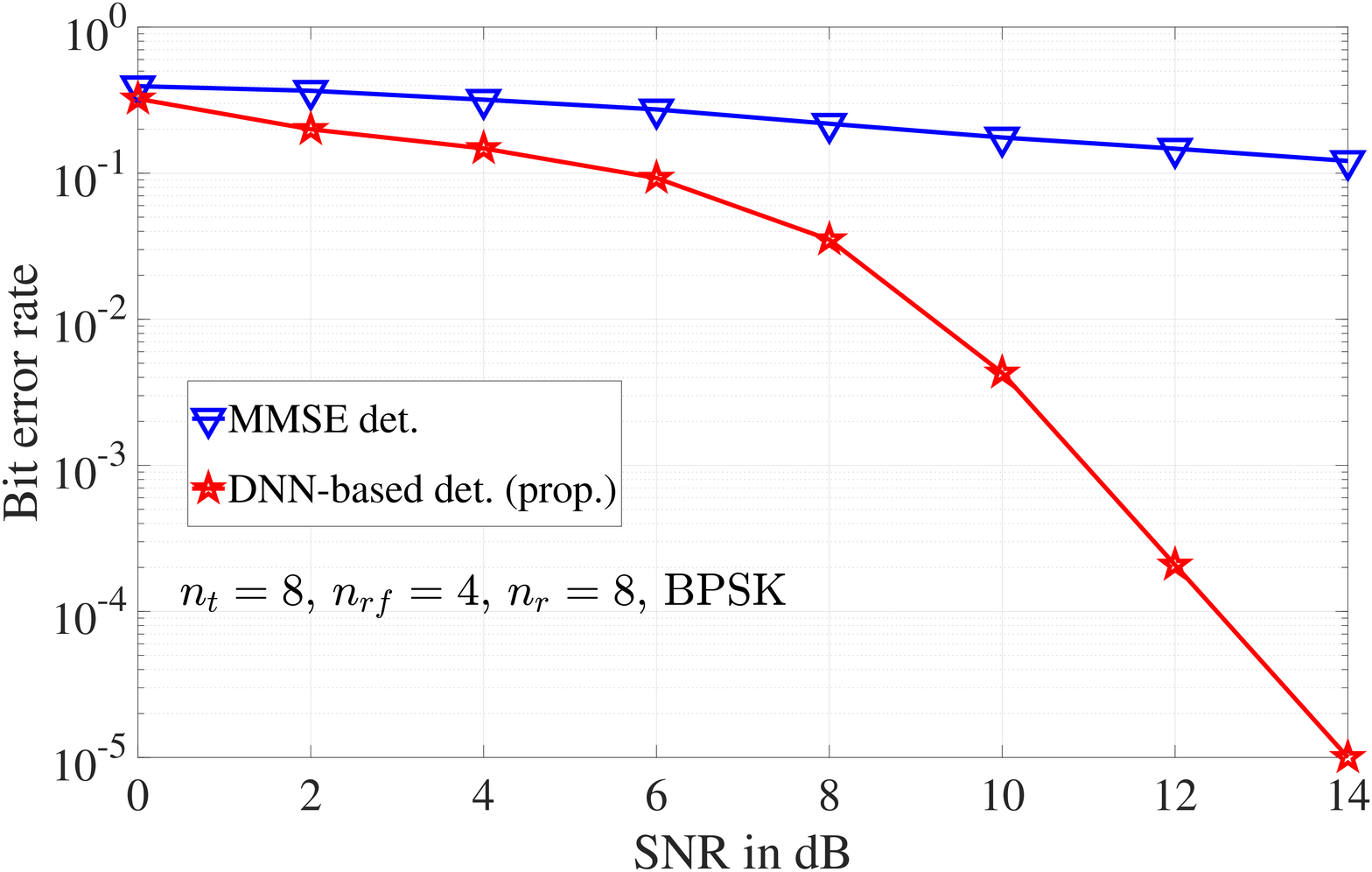}
\caption{}
\label{fig:large_nt8}
\end{subfigure}%
\begin{subfigure}[b]{0.25\textwidth}
\hspace*{-4mm}
\includegraphics[width= 5cm, height=4 cm]{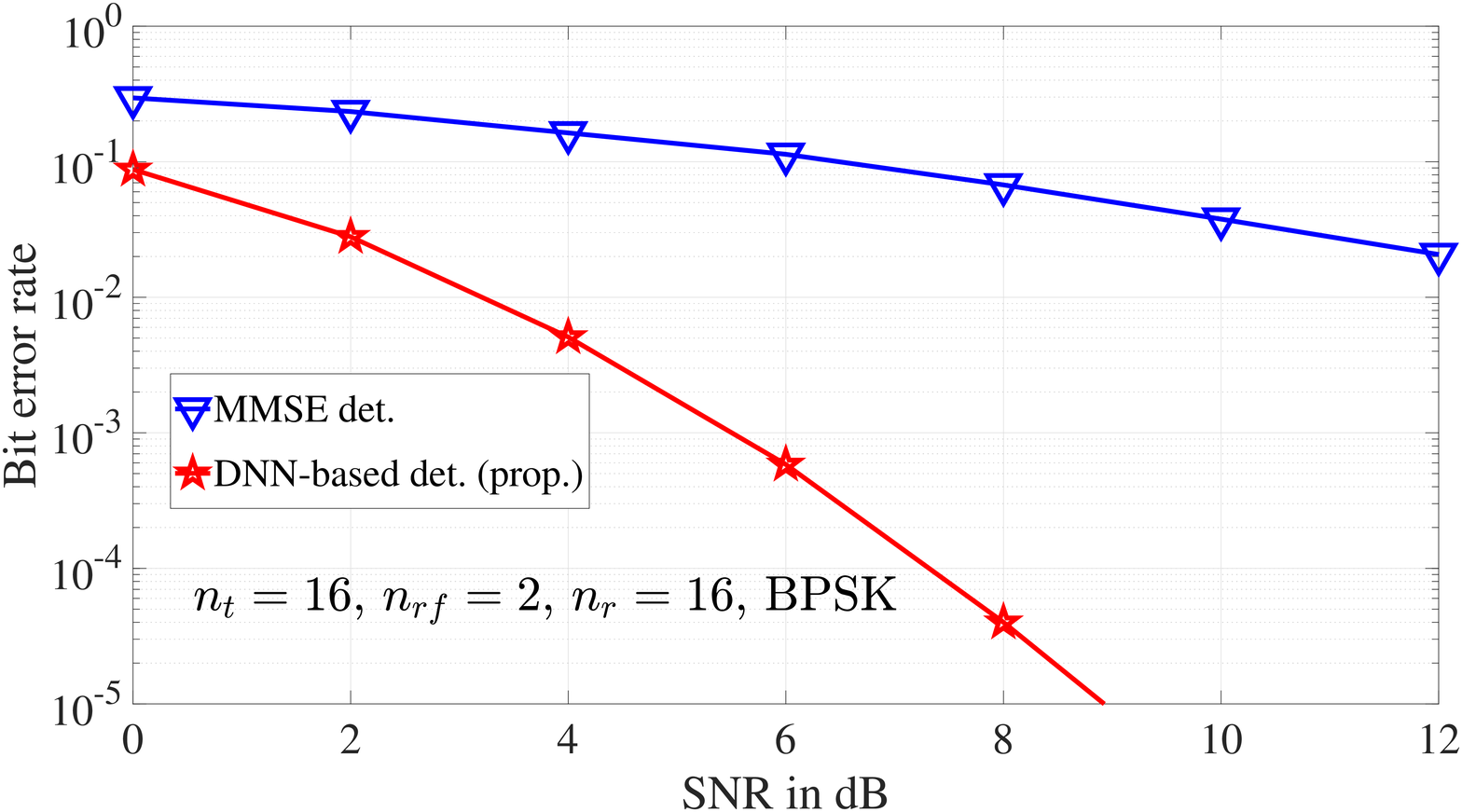}
\caption{}
\label{fig:large_nt16}
\end{subfigure}
\vspace{-4mm}
\caption{ BER performance of large-scale GSM systems with the 
proposed DNN-based detection.
Performance with MMSE detector is also shown for comparison.}
\label{fig:ber_iid_awgn_large}
\vspace{-4mm}
\end{figure}

We next consider large-scale GSM systems which use higher number of 
transmit and receive antennas. Figures \ref{fig:large_nt8} and 
\ref{fig:large_nt16} show the BER performance of two large-scale GSM 
systems using the proposed DNN-based detection. Combinadic encoding 
proposed in \cite{combinadic} is used in both the systems for 
low-complexity encoding of information bits to GSM vectors. The BER 
performance using  MMSE detector is also shown for comparison. The 
DNN parameters used for detection are shown in Table \ref{table2}. 
The following architectures are used: \\ 
{\em 1. GSM system in Fig. \ref{fig:large_nt8}}\\
{\em APP-DNN:} 
$ \text{i/p} \rightarrow 128 \rightarrow \text{ReLU} \rightarrow 64 \rightarrow \text{ReLU} \rightarrow 32 \rightarrow \text{ReLU} \rightarrow 16  \rightarrow \text{ReLU} \rightarrow 8 \rightarrow  \text{sigmoid}$. \\
{\em Symbol-DNN:} 
$\text{i/p} \rightarrow 32 \rightarrow \text{ReLU} \rightarrow 16 \rightarrow \text{ReLU} \rightarrow 8 \rightarrow \text{ReLU} \rightarrow 4  \rightarrow \text{ReLU} \rightarrow 2 \rightarrow  \text{softmax}$. \\
{\em 2. GSM system in Fig. \ref{fig:large_nt16}}\\
{\em APP-DNN:}  
$\text{i/p} \rightarrow 320 \rightarrow \text{ReLU} \rightarrow 160 \rightarrow \text{ReLU} \rightarrow 80 \rightarrow \text{ReLU} \rightarrow 40  \rightarrow \text{ReLU} \rightarrow 20 \rightarrow \text{ReLU} \rightarrow 16 \rightarrow \text{sigmoid}$. \\
{\em Symbol-DNN:} 
$\text{i/p} \rightarrow 128 \rightarrow \text{ReLU} \rightarrow 64 \rightarrow \text{ReLU} \rightarrow 32 \rightarrow \text{ReLU} \rightarrow 16  \rightarrow \text{ReLU} \rightarrow 8 \rightarrow \text{ReLU} \rightarrow 2 \rightarrow  \text{softmax}$. \\
It can be seen from Fig. \ref{fig:ber_iid_awgn_large} that the proposed 
DNN-based detector achieves superior BER performance compared to the 
MMSE detector for both the GSM system settings.

\begin{table}
\centering
\begin{tabular} {|l|c|c|}
\hline 
{\footnotesize Parameters} & {\footnotesize AAP-DNN} & {\footnotesize Symbol-DNN} \\
\hline \hline 
{\footnotesize No. of input neurons} & {\footnotesize Fig. 3a: $2n_r=16$} & {\footnotesize Fig. 3a: $2n_r=16$} \\ 
& {\footnotesize Fig. 3b: $2n_r=32$} & {\footnotesize Fig. 3b: $2n_r=32$} \\ \hline 
{\footnotesize No. of output neurons} & {\footnotesize Fig. 3a: $n_t=8$} & {\footnotesize $|\mathbb{A}|=2$}   \\ 
& {\footnotesize Fig. 3b: $n_t=16$} &    \\ \hline 
{\footnotesize No. of hidden layers} & {\footnotesize Fig. 3a: 4} & {\footnotesize Fig. 3a: 4} \\   
& {\footnotesize Fig. 3b: 6} & {\footnotesize Fig. 3b: 6} \\  \hline 
{\footnotesize Hidden layer activation} & {\footnotesize ReLU} & {\footnotesize ReLU} \\ \hline
{\footnotesize Output layer activation} & {\footnotesize Sigmoid} & {\footnotesize Softmax} \\ \hline
{\footnotesize Optimizer} & {\footnotesize Adam} & {\footnotesize Adam} \\ \hline
{\footnotesize Loss function} & {\footnotesize Binary cross entropy} & {\footnotesize Binary cross entropy} \\ \hline
{\footnotesize No. of training examples} & {\footnotesize 50,000} & {\footnotesize 50,000} \\ \hline
{\footnotesize Training SNR} & {\footnotesize Fig. 3a: 10 dB} & {\footnotesize Fig. 3a: 10 dB} \\ 
& {\footnotesize Fig. 3b: 5 dB} & {\footnotesize Fig. 3b: 5 dB} \\ \hline
{\footnotesize No. of epochs} & {\footnotesize Fig. 3a: 50} & {\footnotesize Fig. 3a: 50}  \\    
& {\footnotesize Fig. 3b: 10} & {\footnotesize Fig. 3b: 10}  \\ 
\hline   
\end{tabular} 
\caption{DNN parameters of the proposed detector for the GSM systems in 
Figs. \ref{fig:large_nt8} and \ref{fig:large_nt16}.} 
\label{table2} 
\end{table}

We compare the complexity (in number of real operations) of the detectors 
considered in Figs. \ref{fig:large_nt8} and \ref{fig:large_nt16} in Table 
\ref{table:complexity}. It can be seen that the proposed detector is 
computationally efficient compared to the exhaustive search based ML 
detection for both the systems. MMSE detector is less complex compared 
to the proposed detector at the cost of degraded BER performance 
(Fig. \ref{fig:ber_iid_awgn_large}). On a machine with Intel i5 (5th gen.) 
processor, the training took less than 30 sec for the system considered 
in Fig. \ref{fig:ber_iid_awgn_small}, and about 2-3 min for the systems 
in Fig. \ref{fig:ber_iid_awgn_large}.

\begin{table}
\centering
\begin{tabular}{|l|c|c|}
\hline
Detector & Complexity for & Complexity for \\
& system in Fig. \ref{fig:large_nt8} & system in Fig. \ref{fig:large_nt16}\\
\hline
ML det. (exhaustive search) & 327680 & 294912 \\
\hline
MMSE det. & 3679 & 25662 \\
\hline
DNN-based det. (prop.) & 34832 & 215876\\
\hline 
\end{tabular}
\caption{Complexity (in number of real operations) of 
algorithms considered in Fig. \ref{fig:ber_iid_awgn_large}.}
\label{table:complexity}
\vspace{-3mm}
\end{table}

\subsection{BER with correlated noise}
Most studies on multiple antenna systems assume the noise 
across the receive antennas to be i.i.d Gaussian. However, this holds 
true only for wide antenna spacing which results in uncoupled antennas. 
In communication devices (e.g., user equipment), there is generally not 
sufficient space to maintain wide-antenna spacing to achieve independent 
noise. Here, we consider the performance of GSM when the noise is correlated 
across different receive antennas \cite{corr_noise},\cite{corr_noise3}. 
A noise correlation matrix characterizing this correlation in multi-antenna 
systems is derived in \cite{corr_noise} by using Nyquist's thermal noise 
theorem. This model depends on the receiver hardware parameters, and hence 
is not a general model for different hardware implementations. 
DNNs are relevant in this context as they can learn to map the 
received signal to the transmitted signal by learning the underlying model 
including the noise correlation specific to the receiver hardware. 
Accordingly, we employ the DNN-based detector proposed in Sec. \ref{Sec2} 
for GSM signal detection in the presence of correlated noise. For the 
purpose of illustration, we consider a correlation model where the  
noise correlation matrix $\mathbf{N}_c$ is of the form 

\begin{equation}
\mathbf{N}_c = \begin{bmatrix}
1 & \rho_n & \rho_n^2 & \dots &\rho_n^{n_r-1} \\
\rho_n & 1 & \rho_n & \dots &\rho_n^{n_r-2}\\
&  & \ddots & &\\
\rho_n^{n_r-1} & \rho_n^{n_r-2} & \dots &  & 1
\end{bmatrix},
\end{equation} 
where $\rho_n \ (0 \leq \rho_n \leq 1)$ is the correlation coefficient. 
With this, the correlated noise across the receive antennas is 
$\mathbf{n}_c = \mathbf{N}_c \mathbf{n}$, where $\mathbf{n}$ is 
i.i.d Gaussian with its entries from $\mathcal{CN}(0, \sigma^2)$.

\begin{figure}[t]
\centering
\includegraphics[width=8cm, height=5.5cm]{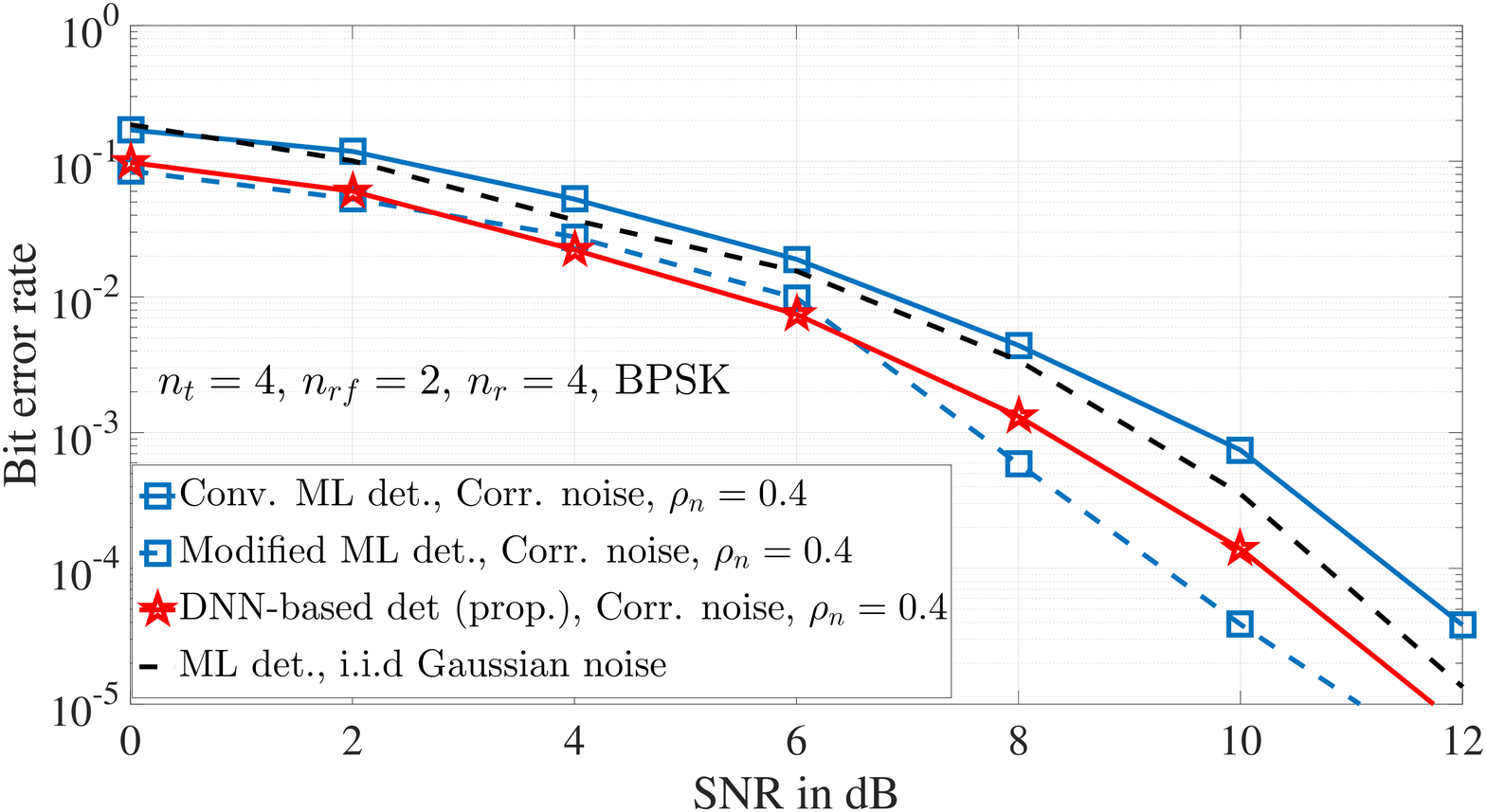}
\caption{BER performance of GSM with the proposed DNN-based detector 
in the presence of correlated noise.}
\vspace{-2mm}
\label{fig:correlated_noise}
\end{figure}

Figure \ref{fig:correlated_noise} shows the BER performance of GSM in the 
presence of correlated noise,  with $\rho_n=0.4$, 
when the proposed DNN-based detector is used. The GSM system and DNN 
architecture parameters considered are the same as those considered in 
Fig. \ref{fig:ber_iid_awgn_small}. 
The performance with the  conventional ML detection in 
\eqref{eq:MLdet} with correlated noise is also shown. Note that 
\eqref{eq:MLdet} is optimum only when noise is i.i.d Gaussian. 
 Therefore, we have also shown the performance  with 
modified ML detection, in which the noise correlation matrix 
$\mathbf{\Sigma}$ is estimated using the procedure in \cite{corr_est} 
and the following modified (optimal) ML detection rule is used:
\begin{equation}
\hat{\mathbf{x}} = \argmin_{\mathbf{x\in \mathbb{S}}} (\mathbf{y} - \mathbf{Hx})^H\mathbf{\Sigma}^{-1}(\mathbf{y} - \mathbf{Hx}).
\label{eq:modified_ML}
\end{equation}
The BER performance of ML detection in \eqref{eq:MLdet} with i.i.d 
Gaussian noise is also shown in the figure for comparison.

The following observations can be made from Fig. \ref{fig:correlated_noise}.
First, it can be seen that the performance of GSM using the conventional 
ML detector in the presence of correlated noise degrades compared to the 
case with i.i.d Gaussian. This is expected because the ML detector in 
\eqref{eq:MLdet} is optimal when the noise across the receive antennas 
is i.i.d Gaussian, and using this detector in correlated noise leads 
to suboptimal detection. Whereas, the performance with the proposed 
DNN-based detector is better than that with the  conventional ML detector 
and is very close to the performance with the modified (optimal) ML 
detector in (\ref{eq:modified_ML}). Further, the performance of the 
proposed detector (and the modified ML detector) with correlated noise 
is better than that of ML detection with i.i.d Gaussian noise. These 
observations can be explained as follows. Among all the noise sequences 
of equal average energy, i.i.d Gaussian noise (uncorrelated noise) is the 
worst-case noise as it has the maximum entropy \cite{shannon},\cite{cover}. 
The correlation introduces a structure in the noise, which allows the DNN 
to learn the model effectively and thus achieve superior performance 
compared to the case of uncorrelated noise.  Although the ML detector 
can be modified to achieve optimal performance as in \eqref{eq:modified_ML}, 
it is limited to the small-scale GSM systems because of its high complexity. 
Whereas the proposed DNN-based detector can be employed in correlated 
noise without any modifications in the architecture. 

\subsection{BER with non-Gaussian noise}
We next consider the case when the noise samples across receive antennas 
are i.i.d, but deviate from Gaussian distribution. Specifically, we 
consider the case when the noise samples have $t$-distribution, 
parameterized by parameter $\nu$. The  $t$-distribution deviates more 
from the Gaussian pdf for smaller values of $\nu$. 

\begin{figure}
\centering
\includegraphics[width=8cm, height=5.5cm]{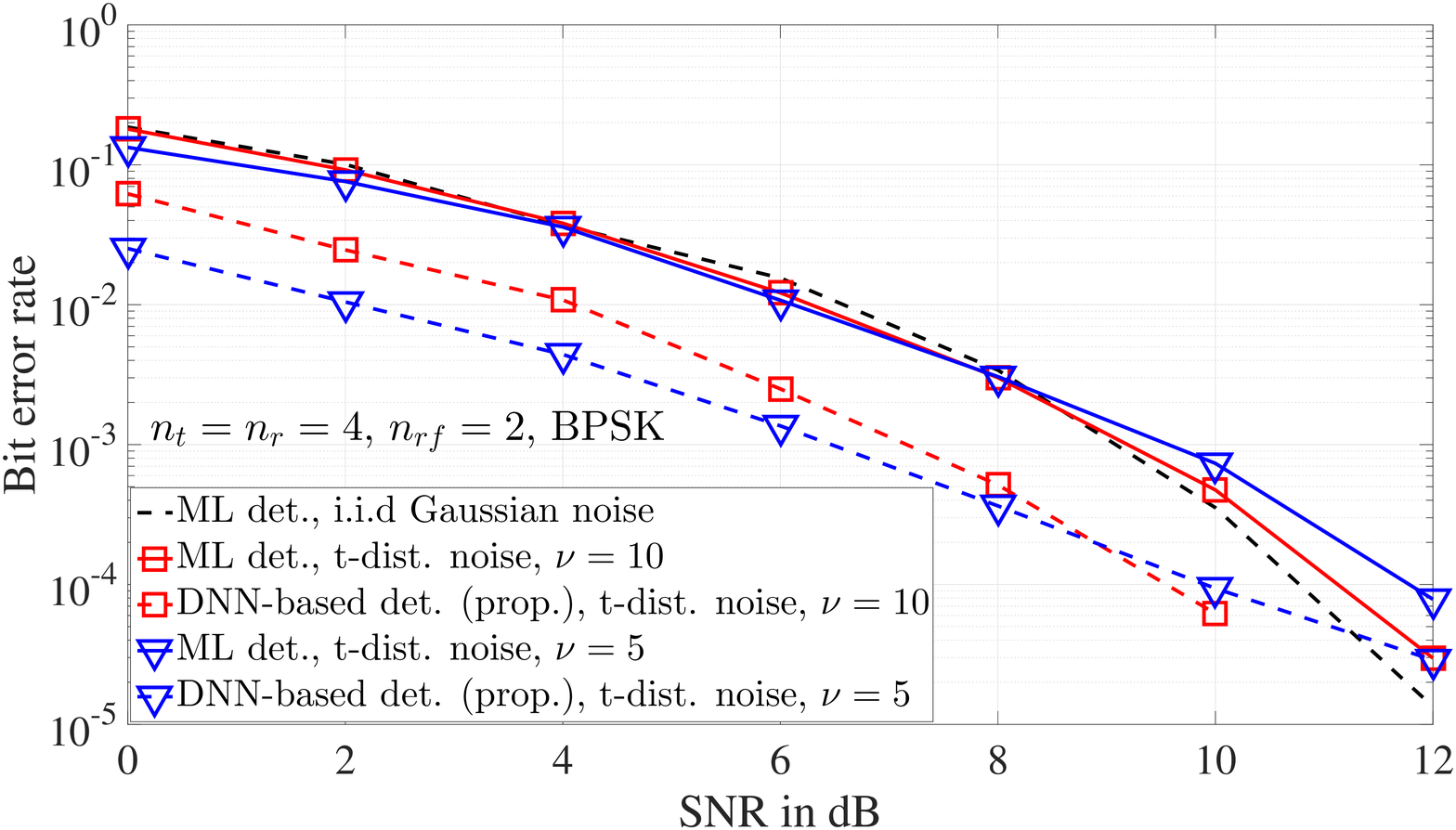}
\caption{BER performance of GSM using the proposed DNN-based detector 
with $t$-distributed noise. 
\vspace{-2mm}}
\label{fig:non-awgn_noise}
\end{figure}

Figure \ref{fig:non-awgn_noise} shows the BER performance of GSM using 
the proposed DNN-based detector when the noise samples are i.i.d across 
the receive antennas and follow $t$-distribution with $\nu=10$ and $\nu=5$. 
The considered GSM system and the DNN architecture are the same as those 
considered in Fig. \ref{fig:ber_iid_awgn_small}. The performance with ML 
detection in \eqref{eq:MLdet} under i.i.d $t$-distributed noise as well as
i.i.d Gaussian noise are also shown for comparison. From Fig. 
\ref{fig:non-awgn_noise}, it can be seen that the performance of ML 
detector with i.i.d $t$-distributed noise is inferior compared to that with 
i.i.d Gaussian noise. It can also be seen that smaller the value of 
$\nu$ (more deviation from Gaussian), more is the degradation in the 
performance of the ML detector. This is mainly because the ML detector in 
\eqref{eq:MLdet} is optimal when the noise is Gaussian distributed and hence 
deviation from Gaussian distribution results in performance degradation. 
Whereas, the proposed DNN based detector shows improved BER performance 
in $t$-distributed noise compared to that in Gaussian noise. As discussed 
earlier, the Gaussian noise is the worst-case noise among all the noise 
distributions for a given variance. Therefore, learning in a non-Gaussian 
noise is more effective, which leads to superior BER performance.  

\subsection{Extension to varying channels}
\begin{figure}
\centering
\includegraphics[width=8cm, height=5.5cm]{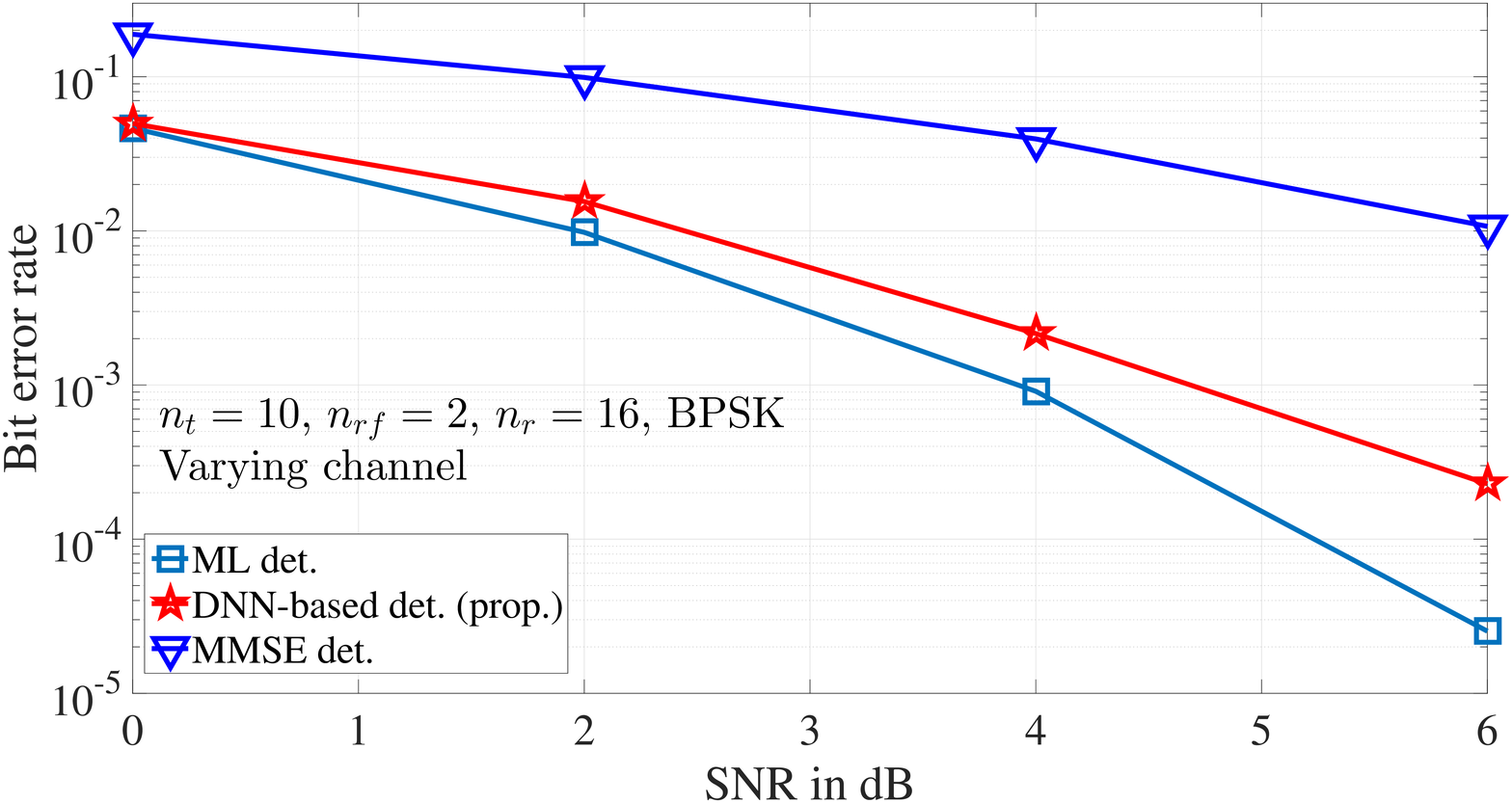}
\caption{BER performance of GSM using the proposed DNN-based detector 
in varying channel.}
\vspace{-2mm}
\label{fig:varying_channel}
\end{figure}

It has been shown in \cite{DetNet} that, using a preprocessing on the 
received vector before feeding it to the DNN can reduce the input 
dimensions, and can enable the DNN to achieve signal detection in varying 
channels (VC). Here, we use the MMSE solution as the preprocessing step to 
achieve dimensionality reduction.
We assume that the channel is known at the receiver, but the channel 
realizations change from one instant to the other. We feed 
{\scriptsize $\mathbf{z}=(\mathbf{H}^H\mathbf{H} + \frac{1}{\text{snr}} \mathbf{I}_{n_t})\mathbf{H}^H \mathbf{y}$} 
as the input to the DNN architecture in Fig. \ref{DGSM_blk_diag} during 
both training and testing phases. Figure \ref{fig:varying_channel} shows 
the BER performance of GSM with the proposed detection architecture using 
MMSE preprocessing in VC. The DNN has 3 sub-DNNs and each of them uses 
5 hidden layers with 320, 256, 128, 64, and 32 neurons in layers 1, 2, 3, 4, 
and 5, respectively. It can be seen that the GSM system with the proposed 
architecture achieves good performance in VC using only 5 hidden 
layers, unlike the detector in \cite{DetNet} which uses 30 layers 
irrespective of the input and output dimensions. 

\section{Conclusions}
\label{sec5}
We proposed a novel modularized DNN-based GSM signal detection scheme. 
Due to its inherent ability to effectively learn the underlying noise 
models in practical receivers, the proposed detector achieved robust and 
better BER performance compared to ML detection performance when deviations 
from the standard model are witnessed. There are several potential directions 
for future work. These include the use of convolutional neural networks in 
the proposed modularized architecture, performance of the proposed detector 
architecture under different channel models, and extension of the detector 
architecture for GSM-based massive MIMO systems.

\end{document}